\documentstyle[manuscript,eqsecnum,aps]{revtex}

\begin{document}
%\rightline{KIAS-P01041} \rightline{SNUTP01-032}

%\centerline{\large\bf Quantum Field Dynamics in a Uniform Magnetic
%Field:} \centerline{\large\bf Description using Fields in Oblique
%Phase Space}\vspace{1cm}

%\centerline { Seok Kim$^1$, Choonkyu Lee$^2$} \centerline{ \it
%Department of Physics and Center for Theoretical Physics}
%\centerline{ \it Seoul National University, Seoul 151-742, Korea}

%\centerline{ Kimyeong Lee$^3$} \centerline{\it School of Physics,
%Korea Institute for Advanced Study} \centerline{\it 207-43,
%Cheongryangri-Dong, Dongdaemun-Gu, Seoul 130-012, Korea}

\title{Quantum Field Dynamics in a Uniform Magnetic Field:\\
Description using Fields in Oblique Phase Space}

\author{Seok Kim$^{a}$\footnote{Email address: calaf2@snu.ac.kr}
, Choonkyu Lee$^{a}$\footnote{Email address:
cklee@phya.snu.ac.kr}, and Kimyeong Lee $^{b}$\footnote{Email
address: klee@kias.re.kr}}

\address{$^{a}$ Department of Physics and Center for Theoretical Physics
\\[-2mm]Seoul National University, Seoul 151-742, Korea}
\address{$^{b}$ School of Physics, Korea Institute for Advanced Study
\\[-2mm] Cheongryangri-Dong, Dongdaemun-Gu, Seoul 130-012, Korea}
\maketitle \vskip -0.7cm

\begin{abstract}
We present a simple field transformation which changes the field
arguments from the ordinary position-space coordinates to the
oblique phase-space coordinates that are linear in position and
momentum variables. This is useful in studying quantum field
dynamics in the presence of external uniform magnetic field: here,
the field transformation serves to separate the dynamics within
the given Landau level from that between different Landau levels.
We apply this formalism to both nonrelativistic and relativistic
field theories. In the large external magnetic field our formalism
provides an efficient method for constructing the relevant
lower-dimensional effective field theories with the field degrees
defined only on the lowest Landau level.
\end{abstract}

\vfill

%\hrule
%\noindent $^1$Email address: calaf2@snu.ac.kr \\
%$^2$Email address: cklee@phya.snu.ac.kr\\
%$^3$Email address: klee@kias.re.kr

\newpage

\setcounter{page}{2}

\section{INTRODUCTION}

In quantum field theory, one deals with a set of fields
$\Psi(\vec{r})$ (time coordinate suppressed), satisfying
appropriate commutation or anticommutation relations and obeying
the operator equations of motion derived from a certain action
functional $S[\Psi(\vec{r})]$. In the initial setup, the arguments
of the quantum fields are usually the position-space coordinates
$\vec{r}=(x^{1}, x^{2},\cdots ,x^{d})$. In the system with
translational invariance, however, one often finds more convenient
description in momentum-space fields $\Phi(\vec{p})$, which are
related to $\Psi(\vec{r})$ by the Fourier transform
\begin{equation}\label{fourier}
\Psi(\vec{r})= \int d^{d}\vec{r}\,\langle\vec{r}|\vec{p}\rangle
\Phi(\vec{p})\hspace{0.15cm},\hspace{0.3cm}
(\langle\vec{r}|\vec{p}\rangle =\frac{1}{(2\pi)^{d/2}}
e^{i\vec{p}\cdot\vec{r}}) \hspace{0.15cm}.
\end{equation}
This field transformation sends the position-space expression
$\hat{p}^{i}\Psi(\vec{r}) \equiv -i\frac{\partial}{\partial
x^{i}}\Psi(\vec{r})$ to $p^{i} \Phi(\vec{p})$, and
$-\hat{x}^{i}\Psi(\vec{r}) \equiv -x^{i}\Psi(\vec{r})$ to
$-i\frac{\partial}{\partial p^{i}}\Phi(\vec{p})$. Now the
following question can be posed: depending on nature of the given
system, can one also utilize the fields $\Phi(\vec{\xi})$ with the
arguments $\vec{\xi} \equiv (\xi^{1},\xi^{2},\cdots,\xi^{d})$
related to $\vec{r}$ and $\vec{p}$ in a less trivial manner? The
answer is yes. In this paper, we will specifically consider the
case that $\xi^{i}\hspace{0.1cm} (i=1, \cdots, d)$ are oblique
phase-space variables of the form
\begin{equation}\label{g-ob-pos}
\xi^{i}= \frac{1}{\sqrt{2}}
\left\{(C)_{ij}x^{j}+(D)_{ij}p^{j}\right\}\hspace{0.15cm},
\end{equation}
where $C,D$ are some real constant matrices. This case is relevant
in studying the dynamics of quantum field systems in the presence
of a uniform background magnetic field.

Regarding $\xi^{i}$ (given by the form (\ref{g-ob-pos})) as new
`coordinate' variables, let us write the related conjugate
`momentum' variables as
\begin{equation}\label{g-ob-mom}
\eta^{i}= \frac{1}{\sqrt{2}}
\left\{(E)_{ij}x^{j}+(F)_{ij}p^{j}\right\}\hspace{0.15cm}.
\end{equation}
This linear canonical transformation of  $x^i, p^i$ requires the
four matrices $C,D,E$ and $F$ to satisfy the conditions
\begin{equation}\label{simplectic}
CD^{T}=DC^{T}\hspace{0.15cm},\hspace{0.3cm}
EF^{T}=FE^{T}\hspace{0.15cm},\hspace{0.3cm}
CF^{T}-DE^{T}=2I\hspace{0.15cm}.
\end{equation}
These linear canonical transformations form a $Sp(2N,R)$ group.
Then the corresponding fields $\Phi(\vec{\xi})$ may be introduced
through the relation analogous to Eq.(\ref{fourier}), i.e., by
considering the generalized Fourier transform
\begin{equation}\label{g-fourier}
\Psi(\vec{r})= \int
d^{d}\vec{\xi}\,\langle\vec{r}|\vec{\xi}\rangle
\Phi(\vec{\xi})\hspace{0.15cm},
\end{equation}
where the basis vector $| \vec{\xi}\rangle$, while obeying the
orthogonality and completeness relations
\begin{equation}\label{orth-complete}
\begin{array}{l}
\langle\vec{\xi}'| \vec{\xi}\rangle = \delta^{d}(\vec{\xi}' -
\vec{\xi})\hspace{0.15cm},\\*[0.1cm] \int d^{d}\vec{\xi}\,
|\vec{\xi}\rangle\langle\vec{\xi}| =1 \hspace{0.15cm},
\end{array}
\end{equation}
should further satisfy the equations
\begin{eqnarray}
&&{\textstyle\langle\vec{r}|\hat{\xi}^{i}|\vec{\xi}\rangle\equiv\frac{1}{\sqrt{2}}
\left[(C)_{ij}x^{j}-i(D)_{ij}\frac{\partial}{\partial x^{j}
}\right]
\langle\vec{r}|\vec{\xi}\rangle=\xi^{i}\langle\vec{r}|\vec{\xi}\rangle}\hspace{0.15cm},
\label{xi-repre}\\
&&{\textstyle\langle\vec{r}|\hat{\eta}^{i}|\vec{\xi}\rangle\equiv\frac{1}{\sqrt{2}}\left[(E)_{ij}x^{j}-i(F)_{ij}\frac{\partial}{\partial
x^{j} }\right] \langle\vec{r}|\vec{\xi}\rangle=
\langle\vec{r}|\vec{\xi}\rangle\left(\overleftarrow{\textstyle\!i\frac{\partial}{\partial
\xi^{i}}}\!\right)}\hspace{0.1cm}.\label{eta-repre}
\end{eqnarray}

By the field transformation (\ref{g-fourier}) we will have the
correspondences
\begin{equation}\label{correspond}
\begin{array}{l}
\hat{\xi^{i}}\Psi(\vec{r})\equiv\frac{1}{\sqrt{2}}\left[(C)_{ij}x^{j}-i(D)_{ij}\frac{\partial}{\partial
x^{j}
}\right]\Psi(\vec{r})\hspace{0.15cm}\leftrightarrow\hspace{0.15cm}\xi^{i}\Phi(\vec{\xi}),\\*[0.3cm]
\hat{\eta^{i}}\Psi(\vec{r})\equiv\frac{1}{\sqrt{2}}\left[(E)_{ij}x^{j}-i(F)_{ij}\frac{\partial}{\partial
x^{j}}\right]\Psi(\vec{r})\hspace{0.15cm}\leftrightarrow\hspace{0.15cm}
-i\frac{\partial}{\partial\xi^{i}}\Phi(\vec{\xi})\hspace{0.15cm}.
\end{array}\end{equation}
The action of the system can then be recast into that involving
the fields $\Phi(\vec{\xi})$, which satisfy suitable
(anti-)commutation relations. In other words we can use the
transformation (\ref{g-fourier}) to obtain a quantum field theory
having $\Phi(\vec{\xi})$ as dynamical fields. It is also not
difficult to find the general expression of the kernel
$\langle\vec{r}|\vec{\xi}\rangle$, as shown in Appendix A. When
both $C$ and $D$ are nonsingular matrices, we may for instance
take $E=-(D^{T})^{-1}$ and $F=(C^{T})^{-1}$. Then the kernel has
the expression
\begin{equation}\label{kernel}
\langle\vec{r}|\vec{\xi}\rangle\!=\!\frac{1}{(2\pi)^{d/2}{\left|
\det\!(\!D/\!\sqrt{2})\right|}^{1/2}}\ e^{-\frac{i}{2}
\{x^{i}(D^{-\!1}C)_{i\!j}x^{j}+\xi^{i}(CD^{T})^{-\!1}_{i\!j}\xi^{j}\}
+\sqrt{2}i\,x^{i}(D^{-\!1})_{i\!j}\xi^{j}}.
\end{equation}
The term in the exponent is the generating fuction of the
canonical linear transformation.

Note that, in quantum field theory, it is usually the form of the
quadratic parts in fields $\Psi$ from the Lagrangian that makes a
certain particular choice for the field arguments more convenient
than others. To see the relevance of the above discussion for a
charged matter system placed in a uniform magnetic field
$\vec{B}=B_{0}\hat{z}$, take a nonrelativistic field system
described by a charged matter field $\Psi(\vec{r},t)$. Here, if
$B_{0}$ is large, the dominant part of the quadratic Lagrangian
will read
\begin{equation}
\int\!\!\!
d^{3}\vec{r}\,\Psi^{\dag}\!{\textstyle(\vec{r},t)\!\left[i\frac{\partial}{\partial
t}\!-\!\frac{1}{2m}\!\left(\!-i\partial_{x}\!\!+\!\frac{qB_{0}}{2}y\right)^{\!\!2}\!\!-\!\frac{1}{2m}
\!\left(\!-i\partial_{y}\!\!-\!\frac{qB_{0}}{2}x\right)^{\!\!2}\!\!-\!\frac{1}{2m}\!\left(\!-i\partial_z\right)^{\!\!2}
\right]}\!\Psi(\vec{r},t),\label{sch-pos-lag}
\end{equation}
when the symmetric gauge for the vector potential,
$\vec{A}(\vec{r})=\left(-\frac{B_{0}}{2}y,
\frac{B_{0}}{2}x,0\right)$, is chosen. Assuming ${\mathcal
B}\equiv -qB_{0}>0$, we may then consider instead of
$(x,y,p_{x},p_{y})$ the following variables
\begin{equation}\label{obl-transf}
\begin{array}{c}
\xi_{1}=\frac{1}{\sqrt{{\mathcal{B}}}}(p_{y}+\frac{1}{2}{\mathcal{B}}x)\hspace{0.15cm},\hspace{0.3cm}
\xi_{2}=\frac{1}{\sqrt{{\mathcal{B}}}}(p_{x}+\frac{1}{2}{\mathcal{B}}y)\hspace{0.15cm},\\*[0.1cm]
\eta_{1}=\frac{1}{\sqrt{{\mathcal{B}}}}(p_{x}-\frac{1}{2}{\mathcal{B}}y)\hspace{0.15cm},\hspace{0.3cm}
\eta_{2}=\frac{1}{\sqrt{{\mathcal{B}}}}(p_{y}-\frac{1}{2}{\mathcal{B}}x)\hspace{0.15cm},
\end{array}\end{equation}
and use the related field transformation (\ref{g-fourier}) to
recast the theory, i.e., as that involving the field
$\Phi(\xi_{1},\xi_{2},\xi_{3}\!\!\equiv\!\! z,t)$. Based on the
correspondences (\ref{correspond}), the quadratic piece in
Eq.(\ref{sch-pos-lag}) is transformed into
\begin{equation}\label{sch-obl-lag}
\int\!\! d^{3}\vec{\xi}\,
\Phi^{\!\dag}\!(\vec{\xi},t)\!\left[i\frac{\partial}{\partial
t}-\frac{{\mathcal{B}}}{2m}\left\{(-i\partial_{\xi_{1}})^{2}+\xi_{1}^{2}\right\}-
\frac{1}{2m}(-i\partial_{\xi_{3}})^{2}\right]\!\Phi(\vec{\xi},t)\hspace{0.15cm},
\end{equation}
viz., we now find the differential operator which has no
dependence on $\xi_{2}$ whatsoever and is of a simple harmonic
oscillator form with respect to $\xi_{1}$. The oscillator states
in the $\xi_{1}$-direction are related to distinct Landau
levels[1], while the coordinate $\xi_{2}$ is associated with
\textit{dynamics within each Landau level}. Now the advantage of
using the $\vec{\xi}$-space field operator
$\Phi(\xi_{1},\xi_{2},z,t)$ should be clear --- it allows one to
study quantum field dynamics in terms of physical excitations
pertaining to one Landau level and, separately, those related to
different Landau-level excitations.

There exist extensive literature on nonrelativistic many-body
theory of charged particles in a background magnetic field,
especially in connection with quantum Hall physics[2]. Similar
studies in relativistic field models have also been made from the
very early days of quantum field theory; a more recent study in
this direction includes the investigation on magnetic catalysis of
dynamical symmetry breaking in certain (2+1)- or (3+1)-
relativistic field theories[3-6]. The crucial factor in these
discussions is the dominance of physical modes belonging to the
lowest Landau level when the background magnetic field is
sufficiently strong. In this light, it is useful to consider the
lowest-Landau-level projection of a given theory and in
Refs.[7-10] issues related to such projection have been discussed.
With our field transformation (\ref{g-fourier}), this projection
can be effected rather trivially \textit{for a quantum field
theory}; for the $\vec{\xi}$-space fields $\Phi(\vec{\xi})$, it
corresponds to a normal dimensional reduction arising when one of
the coordinates becomes essentially compact. Hence our formalism
also suggests a simple, systematic way to construct effective
field theory valid in a large background magnetic field. We should
also mention that, within the nonrelativistic field theory
framework or equivalently nonrelativisitc quantum mechanics, the
authors of Refs.[10,7] previously developed the approach analogous
to ours, while working with holomorphic coordinates (defined in
the coherent-state basis). In this paper we start from the field
transformation (\ref{g-fourier}) to obtain directly a description
of the system based on field operators with oblique phase-space
coordinates. This allows us to utilize the canonical quantization
as much as possible, and furthermore we can apply the same idea on
simplifying the dynamics of relativistic field theories. What we
advocate here is that, in the presence of a strong background
magnetic field, the dynamics of quantum field systems are
described most naturally in terms of field operators defined in
oblique phase space.

This paper is organized as follows. Sec.II contains further
elaborations on nonrelativistic quantum field theory in a uniform
magnetic field, based on the explicit formula for the involved
field transformation. In Sec.III we present corresponding
discussions on relativistic quantum field theory, for both spin-0
and spin-1/2 charged matter fields. Especially, in a strong
magnetic field, dimensionally-reduced field theories are shown to
be relevant at the field operator level. We consider the oblique
phase-space coordinate representation of the Feynman propagator
for the Dirac field in Sec.IV, and then apply this result to
compute the QED one-loop effective action[11-13](with an
appropriate treatment on renormalization). In Sec.V we describe
the gauge interaction term between gaugge and matter fields.  In
Sec. VI, we conclude with some remarks. Appendix A is devoted to
finding the explicit form of the kernel
$\langle\vec{r}|\vec{\xi}\rangle$, while, in Appendix B, we
discuss the form of the field transformation when the background
magnetic
field is represented by the Landau-gauge vector potential.  \\

\section{NONRELATIVISTIC FIELD THEORY}

%\addtocounter{chapter}{1} \setcounter{equation}{0}

Consider a nonrelativistic system involving a Schr\"{o}dinger
field $\Psi(\vec{r},t)$ and its hermitian conjugate, with the
Lagrangian given by the form
\begin{equation}\label{sch-pos-int-lag}
\begin{array}{l}
L={\displaystyle\int}\!d^{3}\vec{r}\,\left(\Psi^{\dag}(\vec{r},t){\textstyle\left(i\frac{\partial}{\partial
t}\right)\Psi(\vec{r},t)}{\displaystyle\frac{}{}}\right.\\*[0.3cm]
\hspace{1.3cm}-\left.\Psi^{\!\dag}\!(\vec{r},t)\!\left[\frac{1}{2m}\!
\left(-i\partial_{x}\!-\!\frac{1}{2}{\mathcal{B}}y\right)^{2}\!\!
+\!\frac{1}{2m}\!\left(-i\partial_{y}\!+\!\frac{1}{2}{\mathcal{B}}x\right)^{2}\!\!
+\!\frac{1}{2m}\!\left(-i\partial_{z}\right)^{2}\right]\!\Psi(\vec{r},t)\right)\\*[0.3cm]
\hspace{0.5cm}-{\displaystyle\int}\!
d^{3}\vec{r}\Psi^{\dag}(\vec{r},t)V^{(1)}(\vec{r})\Psi(\vec{r},t)\\*[0.3cm]
\hspace{0.5cm}-\frac{1}{2}\!{\displaystyle\int}\!
d^{3}\vec{r}d^{3}\vec{r}\,'\,
\Psi^{\dag}(\vec{r},t)\Psi^{\dag}(\vec{r}\,',t)V^{(2)}(\vec{r},\vec{r}\,')
\Psi(\vec{r}\,',t)\Psi(\vec{r},t),
\end{array}\end{equation}
where ${\mathcal{B}}(>0)$ is the constant related to the
background magnetic field strength, and $V^{(1)},V^{(2)}$
represent generic one-body and two-body interactions,
respectively. The fields $\Psi,\Psi^{\dag}$ satisfy the equal-time
(anti-)commutation relations, i.e.,
\begin{equation}\label{pos-comm}
[\Psi(\vec{r},t),\!\Psi(\vec{r}\,'\!,t)]_{\mp}\!\!=\!
[\Psi^{\dag}(\vec{r},t),\!\Psi^{\dag}(\vec{r}\,'\!,t)]_{\mp}\!\!=\!0,\
[\Psi(\vec{r},t),\!\Psi^{\dag}(\vec{r}\,'\!,t)]_{\mp}\!\!=\!\delta^{3}\!(\vec{r}-\!\vec{r}\,'\!),
\end{equation}
where the upper(lower) sign refers to bosons(fermions). [Internal
degrees of freedom, suppressed here for the sake of simplicity,
may be included also.]

Now apply the field transformation (\ref{g-fourier}), choosing
oblique coordinates $\xi_{1},\xi_{2}$ (and $\eta_{1},\eta_{2}$) as
given by (\ref{obl-transf}) and $\xi_{3}\equiv z$. Here, from
Eqs.(\ref{kernel}) and (\ref{obl-transf}), we have the kernel
$\langle\vec{r}|\vec{\xi}\rangle$ given by
\begin{equation}\label{landau-kernel}
\langle\vec{r}|\vec{\xi}\rangle=\frac{\sqrt{{\mathcal{B}}}}{2\pi}\,e^{i\sqrt{{\mathcal{B}}}(x\xi_{2}+y\xi_{1})
-i\frac{{\mathcal{B}}}{2}xy-i\xi_{1}\xi_{2}}\,\delta^{3}(\xi_{3}-\!z)\hspace{0.15cm}.
\end{equation}
Then the result of using the transformation (\ref{g-fourier}) with
the expression (\ref{sch-pos-int-lag}) is the $\vec{\xi}$-space
Lagrangian
\begin{equation}\label{sch-obl-int-lag}
\begin{array}{l}
L\!=\!{\displaystyle\int}\!
d^{3}\vec{\xi}\left(i\Phi^{\dag}\!(\vec{\xi},t)\frac{\partial}{\partial
t}\Phi(\vec{\xi},t)\!-\!\Phi^{\dag}\!(\vec{\xi},t)\!
\left[\frac{{\mathcal{B}}}{2m}\{(-i\partial_{\xi_{1}})^{2}\!+\!\xi_{1}^{2}\}
\!+\!\frac{1}{2m}(-i\partial_{\xi_{3}})^{2}\right]\!\Phi(\vec{\xi},t)\!\right)\\*[0.1cm]
\hspace{0.5cm}-{\displaystyle\int}\!
d^{3}\vec{\xi}d^{3}\vec{\xi}\,'\,\Phi^{\dag}(\vec{\xi},t)
\tilde{V}^{(1)}(\vec{\xi},\vec{\xi}')\Phi(\vec{\xi}',t)\\*[0.1cm]
\hspace{0.5cm}-\frac{1}{2}{\displaystyle\int}\!
d^{3}\vec{\xi}d^{3}\vec{\xi}\,'d^{3}\vec{\xi}\,''d^{3}\vec{\xi}\,'''\,\Phi^{\dag}(\vec{\xi},t)\Phi^{\dag}(\vec{\xi}',t)
\tilde{V}^{(2)}(\vec{\xi},\vec{\xi}',\vec{\xi}'',\vec{\xi}'''\!)\Phi(\vec{\xi}'',t)\Phi(\vec{\xi}'''\!,t),
\end{array}
\end{equation}
where
\begin{eqnarray}\label{interaction}
&&\tilde{V}^{(1)}(\vec{\xi},\vec{\xi}')\!=\!\!{\displaystyle\int}\!
d^{3}\vec{r}\,\langle\vec{\xi}|\vec{r} \rangle V^{(1)}
(\vec{r})\langle\vec{r}|\vec{\xi}'\rangle,\\
&&\tilde{V}^{(2)}(\vec{\xi},\vec{\xi}'\!,\vec{\xi}'',\vec{\xi}''')\!=\!\!{\displaystyle\int}\!
d^{3}\vec{r}d^{3}\vec{r}\,'\langle\vec{\xi}|\vec{r}\rangle\langle\vec{\xi}'|\vec{r}\,'
\rangle V^{(2)}
(\vec{r},\vec{r}\,')\langle\vec{r}\,'|\vec{\xi}''\rangle\langle\vec{r}|\vec{\xi}'''\rangle.
\end{eqnarray}
The new fields $\Phi,\Phi^{\dag}$ also satisfy the
(anti-)commutation relations of the standard form, viz.,
\begin{equation}\label{obl-comm}
[\Phi(\vec{\xi},\!t),\!\Phi(\vec{\xi}\,'\!,\!t)]_{\mp}\!=\!
[\Phi^{\dag}\!(\vec{\xi},\!t),\!\Phi^{\dag}\!(\vec{\xi}\,'\!,\!t)]_{\mp}\!=\!0,\hspace{0.1cm}
[\Phi(\vec{\xi},\!t),\!\Phi^{\dag}\!(\vec{\xi}\,'\!,\!t)]_{\mp}\!=\!\delta^{3}(\vec{\xi}\!-\!\vec{\xi}\,').
\end{equation}
Notice that the ${\mathcal{B}}$-dependent part of the Lagrangian
becomes much simpler in the form (\ref{sch-obl-int-lag}). Using
Eq.(\ref{landau-kernel}), the $\vec{\xi}$-space one-body potential
$\tilde{V}^{(1)}(\vec{\xi},\vec{\xi}')$ may be expressed as
\begin{equation}\label{onebody}
\tilde{V}^{(\!1\!)}(\vec{\xi},\vec{\xi}')\!=\!\delta(\xi_{3}\!-\!\xi'_{3})e^{i(\xi_{1}\xi_{2}-\xi'_{1}\xi'_{2})}
\!\int\!\!
dx\,dy\,e^{-i\sqrt{{\mathcal{B}}}[(\xi_{2}-\xi'_{2})x+(\xi_{1}-\xi'_{1})y]}V^{(\!1\!)}(x,y,\xi_{3}).
\end{equation}
Evidently, no irregularity is observed in considering quantum
fields defined in oblique phase space. For a physical system
defined in two spatial dimensions, it suffices to suppress all
references to the coordinate $z$(i.e., $\xi_{3}$) in our
expressions. We also remark that,instead of working with the
oblique phase-space coordinates $\xi_{1}$ and $\xi_{2}$, one may
alternatively utilize holomorphic coordinates defined with the
help of suitable coherent-state basis[10,7]; but,in dealing with
quantum fields at least, our oblique phase-space description
appears to be more direct and has advantage in that an appropriate
physical interpretation relative to the Landau-level structure may
be given to each (real) coordinate appearing.

In view of the appearance of the oscillator-type differential
operator$\frac{{\mathcal{B}}}{2m}\{(\!-i\partial_{\xi_{1}}\!)^{2}$
$+\xi_{1}^{2}\}$ in Eq.(\ref{sch-obl-int-lag}), one may further
contemplate replacing the (3+1)-dimensional field
$\Phi(\xi_{1},\xi_{2},\xi_{3},t)$ by an $\infty$-component
(2+1)-dimensional fields $\Phi_{n}(\xi_{2},\xi_{3},t)$, where the
discrete index n=0, 1, 2... is associated with a specific
oscillator state for the coordinate $\xi_{1}$. This can be
achieved by writing
\begin{equation}\label{obl-landau-transf}
\Phi(\vec{\xi},t)=\sum_{n=0}^{\infty}\chi_{n}(\xi_{1})\Phi_{n}(\xi_{2},\xi_{3},t)\hspace{0.15cm},
\end{equation}
where $\chi_{n}(\xi_{1})$ is the n-th oscillator eigenfunction,
viz,
\begin{equation}\label{sho}
\chi_{n}(\xi_{1})\equiv\langle\xi_{1}|
n\rangle=\frac{(-i)^{n}}{\pi^{1\!/
\!4}}\frac{1}{\sqrt{2^{n}n!}}H_{n}(\xi_{1})e^{-\frac{1}{2}\xi_{1}^{2}}\hspace{0.15cm},
\end{equation}
$H_{n}(\xi_{1})$ denoting the Hermite polynomial. In this way, we
attach the field $\Phi_{n}(\xi_{2},\xi_{3}\!\equiv\!z,t)$ to the
n-th Landau level. The original field $\Psi(\vec{r},t)$ is related
to the tower of fields $\{\Phi_{n}(\xi_{2},z,t)\}$ by the
transformation
\begin{equation}\label{pos-landau-transf}
\Psi(\vec{r},t)=\sum_{n=0}^{\infty}\int_{-\infty}^{\infty}\!d\xi_{2}\,
{\mathcal{W}}^{(n)}(x,y;\xi_{2})\Phi_{n}(\xi_{2},z,t)
\end{equation}
with the kernel
\begin{equation}\label{pos-landau-kernel}
\begin{array}{l}
{\mathcal{W}}^{(n)}(x,y;\xi_{2})\equiv{\displaystyle\int}_{\!\!-\infty}^{\infty}\!\!d\xi_{1}
\langle x,y|\xi_{1},\!\xi_{2}\rangle\langle\xi_{1}|
n\rangle\\*[0.2cm]
\hspace{2.5cm}=\frac{\sqrt{{\mathcal{B}}}}{2\pi}e^{-i\sqrt{{\mathcal{B}}}(\frac{1}{2}\sqrt{{\mathcal{B}}}y-\xi_{2})x}
{\displaystyle\int}_{\!\!-\infty}^{\infty}\!\!d\xi_{1}\,e^{i(\sqrt{{\mathcal{B}}}y-\xi_{2})\xi_{1}}\chi_{n}(\xi_{1})
\hspace{0.1cm},
\end{array}
\end{equation}
where we used Eq.(\ref{landau-kernel}). Since the Fourier
transform of an oscillator eigenfunction reproduces the same
function in the dual space, the expression in
Eq.(\ref{pos-landau-kernel}) can be simplified:
\begin{equation}
\begin{array}{l}
{\mathcal{W}}^{(\!n\!)}(x,y;\xi_{2}\!)\!=\!\sqrt{\frac{{\mathcal{B}}}{2\pi}}
e^{-i\sqrt{{\mathcal{B}}}(\frac{1}{2}\sqrt{{\mathcal{B}}}y-\xi_{2})x}\,i^{n}\chi_{n}(\sqrt{{\mathcal{B}}}y-\!\xi_{2})\\*[0.2cm]
\hspace{2.25cm}=\!\sqrt{\frac{{\mathcal{B}}}{2\pi}}\frac{1}{\pi^{1\!/\!4}}\frac{1}{\sqrt{2^{n}\!n!}}e^{-i\frac{{\mathcal{B}}}{2}xy}
e^{i\sqrt{{\mathcal{B}}}x\xi_2}H_{n}(\!\sqrt{{\mathcal{B}}}y\!-\!\xi_{2}\!)\,e^{-\frac{1}{2}(\sqrt{{\mathcal{B}}}y-\xi_2)^{2}}.
\end{array}
\end{equation}

By the transformation (\ref{obl-landau-transf}) we obtain an
equivalent (2+1)-dimensional quantum field theory description with
the Lagrangian
\begin{equation}\label{sch-landau-int-lag}
\begin{array}{l}
L={\displaystyle\sum_{n=0}^{\infty}\int_{-\infty}^{\infty}}\!\!d\xi_{2}
{\displaystyle\int}_{-\infty}^{\infty}\!\!dz\,\Phi^{\dag}_{n}\!(\xi_{2},z,t)
\left[i\frac{\partial}{\partial t}\!-\!{\mathcal{E}}_{n}\!
-\!\frac{1}{2m} (-i\partial_{z})^{2}
\right]\Phi_{n}\!(\xi_{2},z,t)
\\*[0.1cm]
\hspace{0.4cm}-{\displaystyle\sum_{n=0}^{\infty}
\sum_{l=0}^{\infty} \!\int\!dz\!
\int\!d\xi_{2}d\xi_{2}'}\,\Phi^{\dag}_{n}\!(\xi_{2},z,t)
V^{(\!1\!)}_{nl}(\xi_{2},\xi_{2}',z)\Phi_{n}\!(\xi_{2}',z,t)\\*[0.1cm]
\hspace{0.4cm} -\frac{1}{2}
{\displaystyle\sum_{n,l,r,s}\!\!\int\!dzdz'\!\!
\int\!d\xi_{2}d\xi_{2}'d\xi_{2}'\!'d\xi_{2}{'''}}\,
\Phi^{\dag}_{n}\! (\xi_{2},z,t) \Phi^{\dag}_{l} \!(\xi_{2}',z',t)
V^{(\!2\!)}_{nlrs}
(\xi_{2},\xi_{2}',\xi_{2}'\!'\!,\xi_{2}{'''}\!,z,z')
\\*[0.1cm]
\hspace{3cm}\times\Phi_{r}\!(\xi_{2}'\!',z',t)\Phi_{s}\!(\xi_{2}'\!'\!',z,t)\hspace{0.2cm},
\end{array}
\end{equation}
where ${\mathcal{E}}_{n}=\frac{{\mathcal{B}}}{m}(n+\frac{1}{2})$,
and
\begin{eqnarray}
&&\!V^{(\!1\!)}_{nl}(\xi_{2},\xi_{2}',z)\!=\!\!\int\!dxdy\,
{\mathcal{W}}^{(\!n\!)\ast}(x,y;\xi_{2})V^{(\!1\!)}(\vec{r})
{\mathcal{W}}^{(l)}(x,y;\xi_{2}')\ ,\\ \label{sch-landau-int.1}
&&\!V^{(\!2\!)}_{nlrs}(\xi_{2},\xi_{2}',\xi_{2}'\!',\xi_{2}{'''},z,z')
\!=\!\!\int\!dxdy\!\int\!dx'dy'\,{\mathcal{W}}^{(\!n\!)\ast}(x,y;\xi_{2})
{\mathcal{W}}^{(l)\ast}(x'\!,y';\xi_{2}')\nonumber\\*[0.1cm]
&&\hspace{4cm}\times V^{(2)}(\vec{r},\vec{r}')
{\mathcal{W}}^{(r)}(x'\!,y';\xi_{2}'\!')
{\mathcal{W}}^{(s)}(x,y;\xi_{2}{'''}).\label{sch-landau-int.2}
\end{eqnarray}
According to this Lagrangian, the terms not involving $V^{(1)}$ or
$V^{(2)}$, i.e., the very terms commonly used to define the
unperturbed propagator, become utmostly trivial --- essentially,
that of free propagation in the z-direction only. The
(2+1)-dimensional fields $\Phi_{n}, \Phi_{l}^{\dag}$ satisfy the
equal-time (anti-)commutation relations
\begin{equation}\label{landau-comm}
\begin{array}{l}
[\Phi_{n}(\xi_2,z,t),\Phi_{l}(\xi_{2}',z',t)]_{\mp}\!=\!
[\Phi_{n}^{\dag}(\xi_2,z,t),\Phi_{l}^{\dag}(\xi_{2}',z',t)]_{\mp}\!=\!0\hspace{0.1cm},\\*[0.1cm]
[\Phi_{n}(\xi_2,z,t),\Phi_{l}^{\dag}(\xi_{2}',z',t)]_{\mp}
\!=\!\delta_{nl}\delta(\xi_{2}\!-\!\xi_{2}')\delta(z\!-\!z')\hspace{0.1cm}.
\end{array}
\end{equation}
We remark that this is a bonafide (2+1)-dimensional field theory;
especially, the variable $\xi_{2}(\equiv
\frac{1}{\sqrt{{\mathcal{B}}}}(p_{x}+\frac{1}{2}{\mathcal{B}}y))$
in the quantum fields $\Phi_{n}(\xi_{2},z,t)$ is just like another
spatial coordinate.

Perturbative or nonperturbative studies on the system can be made
on the basis of the Lagrangian (\ref{sch-landau-int-lag}).
Especially, the functional integration technique with the fields
$\{\Phi_{n}(\xi_{2}, z, t)\}$ may be used. If the magnitude of
$\frac{{\mathcal{B}}}{m}$, the Landau gap, is large relative to
the characteristic energy scale in the problem, it can be
interesting to study dynamics of the system restricted to the
lowest Landau-level states. In such case, all interactions
involving the fields other than the lowest Landau-level field
$\Phi_{0}(\xi_{2},z,t)$ may well be ignored. Actually, more
systematic procedure will be to use the standard effective field
theory approach[14,15] here --- \textit{integrate out} all `heavy'
fields, i.e., the fields $\Phi_{n}(\xi_{2},z,t)$ with $n\neq 0$,
from our theory. Note that, in the latter approach, the
Landau-level mixing effects are included also. Explicit physical
application of this approach, to the quantum Hall physics in
particular, will be pursued elsewhere.

A remark is in order.The Lagrangian form in
Eq.(\ref{sch-obl-int-lag}) or Eq.(\ref{sch-landau-int-lag}) is in
no way specific to our particular gauge choice for the potential
describing the background magnetic field. In Appendix B, it is
shown that the same Lagrangians are obtained even if the Landau
gauge is adopted. It is just the kernel
$\langle\vec{r}|\vec{\xi}\rangle$ in the field transformation
(\ref{g-fourier}) that gets altered with a different gauge chosen.
\\

\section{RELATIVISTIC FIELD THEORY}

%\addtocounter{chapter}{1}\setcounter{equation}{0}

For charged relativistic field systems in the presence of a
background magnetic field, using quantum fields defined in the
oblique phase space can again result in great simplification in
the analysis. In the case of a spin-0 field, the situation is not
much different from the nonrelativistic Schr\"{o}dinger field
case. Let the action, governing the dynamics of a complex spin-0
field $\Psi(\vec{r},t)$ and its hermitian conjugate
$\Psi^{\dag}(\vec{r},t)$, be of the form
\begin{equation}\label{kg-pos-int-lag}
\begin{array}{l}
S=\!\!{\displaystyle\int}\!
dtd^{3}\!\vec{r}\left\{\!\left|\partial_{t}\Psi\right|^{2}\!
\!-\!\left|\left(\partial_{x}\!\!-\!\frac{i}{2}{\mathcal{B}}y\!\right)\Psi\right|^{2}\!
\!-\!\left|\left(\partial_{y}\!\!+\!\frac{i}{2}{\mathcal{B}}x\!\right)\Psi\right|^{2}\!
\!-\!|\partial_{z}\Psi|^{2}\!-\!m^{2}|\Psi|^{2}\!\right\}\\*[0.3cm]
\hspace{0.55cm}+S_{1}(\Psi,\Psi^{\dag},\cdots)\\*[0.3cm]
\hspace{0.35cm}\equiv
S_{0}(\Psi,\Psi^{\dag})+S_{1}(\Psi,\Psi^{\dag},\cdots)\hspace{0.15cm},
\end{array}
\end{equation}
where $S_{1}(\Psi,\Psi^{\dag},\cdots)$ may contain
self-interactions for the fields $\Psi,\Psi^{\dag}$ and possibly
couplings with other dynamical fields. To simplify this system,
especially the piece $S_{0}(\Psi,\Psi^{\dag})$, we may immediately
go over to the description using the $\vec{\xi}$-space fields
$\Phi(\vec{\xi},t),\Phi^{\dag}(\vec{\xi},t)$ through the field
transformation (\ref{g-fourier}) with the kernel
$\langle\vec{r}|\vec{\xi}\rangle$ given by
Eq.(\ref{landau-kernel}). Then we can replace $S_{0}$ by (cf. Eq.
(\ref{sch-obl-lag}))
\begin{equation}\label{kg-obl-lag}
S_{0}\!=\!\int
dtd^{3}\!\vec{\xi}\,\left\{\left|\partial_{t}\Phi\right|^{2}
\!-\!{\mathcal{B}}\left(\left|\partial_{\xi_{1}}\!\Phi\right|^{2}
\!+\!\xi_{1}^{2}\!\left|\Phi\right|^{2}\right)
\!-\!\left|\partial_{\xi_{3}}\!\Phi\right|^{2}\!-\!m^{2}\!\left|\Phi\right|^{2}
\right\}\hspace{0.15cm},
\end{equation}
and $S_{1}$ by the corresponding expression. [If desirable, one
can also consider at this stage suitable field transformations for
other dynamical fields together]. The equal-time commutation
relations satisfied by the fields
$\Phi(\vec{\xi},t),\Phi^{\dag}(\vec{\xi},t)$ and their canonical
momenta
$\partial_{t}\Phi^{\dag}(\vec{\xi},t),\partial_{t}\Phi(\vec{\xi},t)$
are of the identical form, with the coordinates $\vec{\xi}$ taking
the place of $\vec{r}$ , as those satisfied by the original
configuration-space fields.

The above $\vec{\xi}$-space description can be the basis of all
field-theoretical investigations on the system, and it will have
advantage, for instance, by having the simpler Feynman propagator
expression (as determined by the form (\ref{kg-obl-lag})). [See
the remark following after Eq.(\ref{kg-prop})]. When the value of
${\mathcal{B}}$ is large, one can also utilize the equivalent
(2+1)-dimensional field-theory description, based on the set of
complex fields $\Phi_{n}(\xi_{2},\xi_{3},t)$, $n=0,1,2,\cdots$,
where $n$ refers to the n-th Landau level. It is achieved by the
field transformation (\ref{pos-landau-transf}). In the latter
description, the action $S_{0}$ will be replaced by
\begin{equation}\label{kg-landau-lag}
S_{0}\!=\!\sum_{n=0}^{\infty}\!\int\!
dtd\xi_{2}d\xi_{3}\left\{|\partial_{t}\Phi_{\!n}(\xi_{2},\xi_{3},t)|^{2}
\!\!-\!|\partial_{\xi_{3}}\!\Phi_{\!n}(\xi_{2},\xi_{3},t)|^{2}
\!\!-\!M_{\!n}^{2}|\Phi_{\!n}(\xi_{2},\xi_{3},t)|^{2}\!\right\},
\end{equation}
where
\begin{equation}\label{kg-eff-mass}
M_{n}=\sqrt{m^{2}+{\mathcal{B}}(2n+1)}\hspace{0.2cm}.
\end{equation}
The fields $\Phi_{n}(\xi_{2},\xi_{3},t)$ will satisfy the standard
equal-time commutation relations of any (2+1)-dimensional
relativistic spin-0 field theory. From the form
(\ref{kg-landau-lag}) we notice that the Feynman propagator
associated with the n-th Landau-level field $\Phi_{n}$ has a
particularly simple structure---aside from the trivial
$\delta(\xi_{2}\!-\!\xi_{2}')$ factor, it is just the
(1+1)-dimensional free scalar propagator for a particle of mass
$M_{n}$. All effective mass values $M_{n}$ increase indefinitely
with ${\mathcal{B}}$; this suggests that if the background
magnetic field becomes sufficiently large, the given system(even
for \textit{relatively small m}) may be described approximately
using the appropriate nonrelativistic field-theory description.

The Dirac field dynamics in a given background magnetic field can
be more interesting. If $\Psi(\vec{r},t)$ now denotes the Dirac
field, the part of the action we denoted above as $S_{0}$ will
have the form
\begin{equation}\label{dirac-pos-lag}
S_{0}\!\!=\!\int\!\!
dtd^{3}\!\vec{r}\,\Psi^{\!\dag}\!(\vec{r},t)i\!\left\{\partial_{t}
\!+\!\alpha_{1}\!\left(\!\partial_{x}\!-\!\frac{i}{2}{\mathcal{B}}y\!\right)
\!+\!\alpha_{2}\!\left(\!\partial_{y}\!+\!\frac{i}{2}{\mathcal{B}}x\!\right)
\!+\!\alpha_{3}\partial_{z}\!+\!i\beta m\right\}\!\Psi(\vec{r},t),
\end{equation}
where, for the four $4\times 4$ matrices $\vec{\alpha}$ and
$\beta$, we may assume the Dirac representation:
\begin{equation}\label{gamma}
\alpha_{i}\equiv\gamma^{0}\gamma^{i}=\left(\begin{array}{cc} 0 &
\sigma_{i} \\ \sigma_{i} & 0
\end{array}\right)\hspace{0.15cm},\hspace{0.3cm}
\beta\equiv\gamma^{0}=\left(\begin{array}{cc} I & 0 \\
0 & -I \end{array}\right)\hspace{0.15cm}.
\end{equation}
The fields $\Psi(\vec{r},t)$ and $\Psi^{\dag}(\vec{r},t)$ satisfy
the equal-time anticommutation relations. For the interaction term
$S_{1}(\Psi,\Psi^{\dag},\cdots)$, the most interesting case will
be the one describing the matter coupling with dynamical gauge
fields[3-6] (or with Chern-Simons-type gauge fields in the
(2+1)-dimensional case). But our principal concern here is the
quadratic action (\ref{dirac-pos-lag}). As the field
transformation (\ref{g-fourier}) is applied to the Dirac fields,
the following $\vec{\xi}$-space action is obtained:
\begin{equation}\label{dirac-obl-lag}
S_{0}\!=\!\int\!\! dtd^{3}\!\vec{\xi}\
\Phi^{\!\dag}\!(\vec{\xi},t)\left\{i\partial_{t}
\!+\!i\alpha_{1}\sqrt{{\mathcal{B}}}\frac{\partial}{\partial\xi_{1}}
\!-\!\alpha_{2}\sqrt{{\mathcal{B}}}\xi_{1}\!+\!i\alpha_{3}\partial_{\xi_{3}}\!-\!\beta
m\right\}\!\Phi(\vec{\xi},t)\hspace{0.1cm}.
\end{equation}
These $\vec{\xi}$-space fields satisfy the standard equal-time
anticommutation relations also. For discussions somewhat similar
to our treatment below(but without utilizing oblique phase-space
coordinates), see Ref.[6].

Now, from the differential operator appearing in
Eq.(\ref{dirac-obl-lag}), we observe that the piece involving
$\xi_{1}$ can be written as
\begin{equation}\label{sho-structure}
\begin{array}{ll}
i\alpha_{1}\sqrt{{\mathcal{B}}}\frac{\partial}{\partial\xi_{1}}
\!-\!\alpha_{2}\sqrt{{\mathcal{B}}}\xi_{1}=
\sqrt{{\mathcal{B}}}\alpha_{2}\left(\xi_{1}\!-\!\Sigma_{3}
\frac{\partial}{\partial\xi_{1}}\right)&\hspace*{0.5cm}\\
\hspace*{3.3cm}=-\sqrt{2{\mathcal{B}}}\alpha_{2}
\left(\!\begin{array}{cccc}
\ \bar{a}\ &\ 0\ &\ 0\ &\ 0\ \\
\ 0\ &\ a\ &\ 0\ &\ 0\ \\
\ 0\ &\ 0\ &\ \bar{a}\ &\ 0\ \\
\ 0\ &\ 0\ &\ 0\ &\ a\ \end{array}\!\right),
\end{array}
\end{equation}
where we have defined
$a=\frac{1}{\sqrt{2}}\left(\xi_{1}+\frac{\partial}{\partial\xi_{1}}\right)$,
$\bar{a}=\frac{1}{\sqrt{2}}\left(\xi_{1}-\frac{\partial}{\partial\xi_{1}}\right)$,
and $\frac{1}{2}\Sigma_{3}$ is the third component of the spin
angular momentum
\begin{equation}\label{spin-matrix}
\Sigma_{3}=-i\alpha_{1}\alpha_{2}=\left(
\begin{array}{cc}\sigma_{3}&0\\0&\sigma_{3}\end{array}\right)\hspace{0.1cm}.
\end{equation}
With this understanding, we may then introduce the following
transformation on the fields $\Phi(\vec{\xi},t)$,
$\Phi^{\dag}(\vec{\xi},t)$:
\begin{equation}\label{dirac-obl-landau-transf}
\left(\Phi(\vec{\xi},t)\right)_{\alpha}
=\sum_{n=0}^{\infty}\left({\mathcal{M}}_{n}(\xi_{1})\right)_{\alpha\beta}
\left(\tilde{\Phi}_{n}(\xi_{2},\xi_{3},t)\right)_{\beta}\hspace{0.15cm},
\end{equation}
\begin{equation}\label{sho-matrix}
{\mathcal{M}}_{n}(\xi_{1})=\left(\!\begin{array}{cccc}
\chi_{\!n\!-\!1}\!(\xi_{1}\!)&0&0&0\\
0&\chi_{\!n}\!(\xi_{1}\!)&0&0\\
0&0&\chi_{\!n\!-\!1}\!(\xi_{1}\!)&0\\
0&0&0&\chi_{\!n}\!(\xi_{1}\!)
\end{array}\!\right),
\end{equation}
where $\chi_{n}(\xi_{1})$ $(n=0, 1, 2, \cdots)$ denote the
harmonic oscillator wave functions in Eq.(\ref{sho}), and
$\chi_{-1}(\xi_{1})\equiv 0$. The index $n$ again refers to a
specific Landau level, and our (2+1)-dimensional spinor fields
$\tilde{\Phi}_{n}(\xi_{2},\xi_{3},t)$ for $n\neq 0$ have four
independent components while the lowest Landau-level field,
$\tilde{\Phi}_{0}(\xi_{2},\xi_{3},t)$, has two independent
components (with $\Sigma_{3}'=-1$) only, i.e.,
\begin{equation}\label{grd-landau-field}
\left(\tilde{\Phi}_{0}(\xi_{2},\!\xi_{3},\!t)\right)_{\alpha}
=\left(\begin{array}{c} 0\\
\left(\!\tilde{\Phi}_{0}(\xi_{2},\!\xi_{3},\!t)\right)_{2}
\\0\\ \left(\tilde{\Phi}_{0}(\xi_{2},\!\xi_{3},\!t)\right)_{4}
\end{array}\!\right)\hspace{0.1cm}.
\end{equation}
Using these (2+1)-dimensional spinor fields, the action
(\ref{dirac-obl-lag}) can be recast as
\begin{equation}\label{dirac-landau-lag}
S_{0}=\!\sum_{n=0}^{\infty}\!\int\!
dtd\xi_{2}d\xi_{3}\,\tilde{\Phi}_{n}^{\!\dag}\!(\xi_{2},\!\xi_{3},\!t)\left\{i\partial_{t}
\!+\!i\alpha_{3}\partial_{\xi_{3}}\!\!-\!\left(m\beta\!+\!\sqrt{2{\mathcal{B}}n}\,\alpha_{2}\right)
\!\right\}\tilde{\Phi}_{n}(\xi_{2},\!\xi_{3},\!t).
\end{equation}
Observe that, using the two-component field
$\tilde{\Phi}_{0}=\left(\!\begin{array}{c}(\tilde{\Phi}_{0})_{2}\\(\tilde{\Phi}_{0})_{4}\end{array}\!\right)$,
the differential operator appearing in the $n=0$ term of this
expression may be written
$\{i\partial_{t}-i\sigma_{1}\partial_{\xi_{3}}-m\sigma_{3}\}$.

As we have seen above, the given Dirac field system can be
reformulated as a (2+1)-dimensional field theory in oblique phase
space. The action (\ref{dirac-landau-lag}) can be changed to have
more standard form with a diagonal mass matrix, by redefining our
fields according to
\begin{equation}\label{landau-eff-transf}
\left(\tilde{\Phi}_{n}(\xi_{2},\xi_{3},t)\right)_{\alpha}
=\left(U_{n}\right)_{\alpha\beta}
\left(\Phi_{n}(\xi_{2},\xi_{3},t)\right)_{\beta},
\end{equation}
where
\begin{equation}\label{eff-transf-matrix}
U_{n}\!=\!\left(\!\begin{array}{cccc}
\cos\!\frac{\theta_{n}}{2}&0&0&i\!\sin\!\frac{\theta_{n}}{2}\\
0&\cos\!\frac{\theta_{n}}{2}&-i\!\sin\!\frac{\theta_{n}}{2}&0\\
0&-i\!\sin\!\frac{\theta_{n}}{2}&\cos\!\frac{\theta_{n}}{2}&0\\
i\!\sin\!\frac{\theta_{n}}{2}&0&0&\cos\!\frac{\theta_{n}}{2}
\end{array}\!\right),\hspace{0.15cm}
{\textstyle\left(\tan\theta_{n}=\frac{\sqrt{2{\mathcal{B}}n}}{m}\right)}.
\end{equation}
The matrix $U_{n}$ is unitary and has the properties
\begin{equation}\label{eff-gamma-transf}
\begin{array}{l}
U_{n}^{-1}(m\beta+\sqrt{2{\mathcal{B}}n}\,\alpha_{2})U_{n}=\sqrt{m^{2}\!+\!2{\mathcal{B}}n}\,\beta
\hspace{0.1cm},\\*[0.15cm]
U_{n}^{-1}\alpha_{3}U_{n}=\alpha_{3}\hspace{0.1cm}.
\end{array}
\end{equation}
Thus, in terms of the spinor fields $\Phi_{n}(\xi_{2},\xi_{3},t)$,
the action (\ref{dirac-landau-lag}) assumes the form
\begin{equation}\label{dirac-eff-lag}
S_{0}=\sum_{n=0}^{\infty}\int\!
dtd\xi_{2}d\xi_{3}\,\Phi^{\dag}_{n}(\xi_{2},\!\xi_{3},\!t)\left\{i\partial_{t}
\!+\!i\alpha_{3}\partial_{\xi_{3}}\!-\!M_{n}\beta
\right\}\Phi_{n}(\xi_{2},\!\xi_{3},\!t)\hspace{0.1cm},
\end{equation}
where $M_{n}=\sqrt{m^{2}+2{\mathcal{B}}n}$ denotes the mass
parameter associated with the (2+1)-dimensional spinor field
$\Phi_{n}(\xi_{2},\xi_{3},t)$. The fields
$\Phi_{n}(\xi_{2},\xi_{3},t)$,
$\Phi^{\dag}_{n}(\xi_{2},\xi_{3},t)$ also satisfy the standard
equal-time anticommutation relations. {}From
Eqs.(\ref{dirac-obl-landau-transf}) and (\ref{landau-eff-transf})
the transformation relating the original Dirac field
$\Psi(\vec{r},t)$ to the set of (2+1)-dimensional fields
$\{\Phi_{n}(\xi_{2},\xi_{3},t)\}$, i.e., the formula involving the
functions ${\mathcal{W}}^{(n)}(x,y;\xi_{2})$ as in
Eq.(\ref{pos-landau-transf}), can be found as well.

When only the quadratic action in Eq.(\ref{dirac-eff-lag}) is
taken into account, we notice that dynamics of the field
$\Phi_{n}(\xi_{2},\xi_{3},t)$ is simply that of a free
\textit{(1+1)-dimensional spinor field theory}, with the
coordinate $\xi_{2}$, which is supposed to describe dynamics
within a given Landau level, playing a rather trivial role. This
situation is similar to the case of the spin-0 field systems
discussed earlier. But there is one, physically significant,
difference. In this Dirac field system, the effective mass
parameter for the n-th Landau-level field $\Phi_{n}$ equals
$\sqrt{m^{2}+2{\mathcal{B}}n}$ and hence the associated mass for
the lowest Landau level field, i.e., the two-component field
$\Phi_{0}(\xi_{2},\xi_{3},t)$(=$\tilde{\Phi}_{0}(\xi_{2},\xi_{3},t)$),
is equal to m regardless of the magnitude of ${\mathcal{B}}$. Now
consider a field theory system with the action given by the
quadratic action (\ref{dirac-pos-lag}) plus some nontrivial
interaction $S_{1}(\Psi,\Psi^{\dag},\cdots)$ (which may also be
expressed using the fields $\Phi_{n}(\xi_{2},\xi_{3},t),\
n\!=\!0,1,2,\cdots$). In a sufficiently large background magnetic
field, one may then integrate out all ``heavy'' fields
$\Phi_{n}(\xi_{2},\xi_{3},t),\ n=1,2,\cdots\ $ from the system and
go on to discuss various physical problems in terms of the
resulting effective field theory which involves just the lowest
Landau-level field $\Phi_{0}(\xi_{2},\xi_{3},t)$. This scheme
should have a useful application in studying, for instance, the
magnetic catalysis of chiral symmetry breaking[3-6] in a system
with a vanishingly small mass m.
\\

\section{ FEYNMAN PROPAGATOR AND THE ONE-LOOP EFFECTIVE
ACTION}

%\addtocounter{chapter}{1} \setcounter{equation}{0}

For the Dirac field system discussed in the previous section, we
shall here find the Feynman propagator associated with the
$\vec{\xi}$-space fields, i.e.,
\begin{equation}\label{dirac-prop-def}
S_{\alpha\beta}(\vec{\xi},t;\vec{\xi}\,',t') \!=\!-i\langle
0|T(\Phi_{\alpha}(\vec{\xi},t)
\bar{\Phi}_{\beta}(\vec{\xi}\,',t')\,)|0\rangle,\hspace{0.15cm}
\left(\bar{\Phi}(\vec{\xi},t)\equiv\Phi^{\dag}(\vec{\xi},t)\gamma^{0}\right)
\end{equation}
and also calculate in this $\vec{\xi}$-space framework the
one-loop effective action explicitly. For these considerations, it
is the $\vec{\xi}$-space spinor field action (\ref{dirac-obl-lag})
that is relevant. We do this analysis not only to exhibit the
simplicity of the $\vec{\xi}$-space description but also to show
that all standard field theoretic manipulations, including
renormalization, may be carried out in this description.

The propagator defined above satisfies the differential equation
\begin{equation}\label{green-ftn-eq}
{\textstyle(i\gamma^{0}\partial_{t}
\!+\!i\sqrt{{\mathcal{B}}}\gamma^{1}\partial_{\xi_{1}}
\!+\!i\gamma^{3}\partial_{\xi_{3}}\!\!-\!\gamma^{2}\sqrt{{\mathcal{B}}}\xi_{1}
\!-\!m)_{\alpha\!\gamma}S_{\gamma\!\beta}(\vec{\xi},t;\vec{\xi}',t')
\!=\!\delta_{\alpha\beta}\delta^{3}(\vec{\xi}\!-\!\vec{\xi}')\delta(t\!-\!t')}.
\end{equation}
Introducing the shorthand notations
\begin{equation}
\begin{array}{l}
i\gamma^{0}\partial_{t}
\!+\!i\sqrt{{\mathcal{B}}}\gamma^{1}\partial_{\xi_{1}}
\!+\!i\gamma^{3}\partial_{\xi_{3}}\equiv
i\tilde{\partial}\hspace{-0.19cm}\slash\hspace{0.15cm},\\*[0.2cm]
(i\tilde{\partial}\hspace{-0.19cm}\slash)^{2}=-(\partial_{t}^{2}
\!-\!{\mathcal{B}}\partial_{\xi_{1}}^{2}\!-\!\partial_{\xi_{3}}^{2})
\equiv-\tilde{\partial}^{2}
\end{array}
\end{equation}
and employing the matrix notation in the Dirac spin space, we may
then write
\begin{equation}\label{dirac-kg-relation}
\begin{array}{l}
S(\xi;\xi')\!=\!\langle\xi|
{\displaystyle\frac{1}{i\tilde{\partial}\hspace{-0.19cm}\slash-\!\gamma^{2}\sqrt{{\mathcal{B}}}\xi_{1}
\!-\!m\!+\!i\epsilon}|\xi'\rangle}\\*[0.35cm]
\hspace{1.15cm}=(i\tilde{\partial}\hspace{-0.19cm}\slash-\!\gamma^{2}\sqrt{{\mathcal{B}}}\xi_{1}\!+\!m)
\langle\xi|{\displaystyle\frac{1}{-\tilde{\partial}^{2}\!-\!{\mathcal{B}}\xi_{1}^{2}
\!-\!{\mathcal{B}}\Sigma_{3}\!-\!m^{2}\!+\!i\epsilon}}|\xi'\rangle\\*[0.37cm]
\hspace{1.15cm}\equiv
(i\tilde{\partial}\hspace{-0.19cm}\slash-\!\gamma^{2}\sqrt{{\mathcal{B}}}\xi_{1}\!+\!m)
G(\xi;\xi')\hspace{0.15cm},
\end{array}
\end{equation}
where we have used the relation
\begin{equation}
(i\tilde{\partial}\hspace{-0.19cm}\slash-\!\gamma^{2}\sqrt{{\mathcal{B}}}\xi_{1}
\!-\!m)(i\tilde{\partial}\hspace{-0.19cm}\slash-\!\gamma^{2}\sqrt{{\mathcal{B}}}\xi_{1}
\!+\!m)\!=\!-\tilde{\partial}^{2}\!-\!{\mathcal{B}}\xi_{1}^{2}
\!-\!{\mathcal{B}}\Sigma_{3}\!-\!m^{2}\hspace{0.15cm}.
\end{equation}
For the quadratic propagator $G(\xi;\xi')$ it is advantageous to
use the Schwinger proper-time representation[12], that is,
\begin{eqnarray}
&&G(\xi,\xi')\!=\!\int_{0}^{\infty}\!ds\,e^{-\epsilon s}
\langle\xi s|\xi'\rangle\ ,\label{schwinger-para}\\ & &\langle\xi
s|\xi'\rangle\!=\!
\langle\xi|e^{-is(\tilde{\partial}^{2}+{\mathcal{B}}\xi_{1}^{2}
+{\mathcal{B}}\Sigma_{3}+m^{2})}|\xi'\rangle\ ,\label{trans-ampl}
\end{eqnarray}
and $\langle\xi s|\xi'\rangle\ $ can be found from the well-known
expression for the free or harmonic oscillator propagator. In this
way, we immediately obtain
\begin{equation}\label{kg-prop}
\begin{array}{l}
G(\xi,\xi')=\delta(\xi_{2}\!-\!\xi_{2}')\!{\displaystyle\int}_{
\!0}^{\infty}
\!ds\,e^{-is(m^{2}+{\mathcal{B}}\Sigma_{3}-i\epsilon)}
\!\sqrt{\!\frac{i}{4\pi s}}\!\sqrt{\!\frac{1}{4\pi is}}
e^{-i\frac{(t-t'\!)^{\!2}}{4s}\!+\!i\frac{(z-z'\!)^{\!2}}{4s}}\\*[0.2cm]
\hspace{4cm}\times\sqrt{\frac{1}{2\pi i\sin(2{\mathcal{B}}s)}}
\,e^{\frac{i}{2}\frac{(\xi_{1}^{2}+{\xi_{1}'}^{\!2})\!\cos(\!2{\mathcal{B}}s\!)
-2\xi_{1}\xi_{1}'}{\sin(2{\mathcal{B}}s)}}.
\end{array}
\end{equation}
Substitution of the form (\ref{kg-prop}) into
Eq.(\ref{dirac-kg-relation}) gives the desired expression for
$S(\xi;\xi')$. [Incidentally, from the integral representation
(\ref{kg-prop}), suppressing the factor
$e^{-is{\mathcal{B}}\Sigma_{3}}$ inside the integrand yields the
related Feynman propagator for the spin-0 field].

The above information may be used to calculate the one-loop spinor
effective action in a uniform magnetic field. In position space,
the (unenormalized) one-loop effective action can be represented
by the proper-time integral[12,13]
\begin{equation}\label{eff-act-pos-integ}
\Gamma^{(1\!)}\!({\mathcal{B}})=\frac{i}{2}\!\int_{0}^{\infty}\!\frac{ds}{s}
e^{-\epsilon s}\int d^{4}x\,\lim_{x'\rightarrow x} {\rm tr}
\langle\,xs| x'\rangle\hspace{0.15cm},
\end{equation}
where $\langle\,xs| x'\rangle$ is the proper-time Green's function
associated with the `quadratic' Dirac operator. But the function
$\langle\,xs| x'\rangle$ is related to $\langle\,\xi s|
\xi'\rangle$ (see Eq.(\ref{trans-ampl})) by
\begin{equation}
\langle\,xs| x'\rangle=\int d^{4}\xi d^{4}\xi'\langle\,x|
\xi\rangle \langle\xi s|\xi'\rangle\langle\xi'|x'\rangle
\end{equation}
(with $d^{4}\xi\equiv dt\,d^{3}\vec{\xi}$ and $\langle\,x|
\xi\rangle=\delta(x^{0}-\xi^{0})\langle\vec{r}|\vec{\xi}\rangle$),
and hence, formally,
\begin{equation}\label{eff-act-obl-integ}
\int d^{4}x\,\lim_{x'\rightarrow x} {\rm tr} \langle\,xs|
x'\rangle =\int d^{4}\xi\,\lim_{\xi'\rightarrow \xi} {\rm tr}
\langle\,\xi s| \xi'\rangle\hspace{0.15cm}.
\end{equation}
Using the explicit expression for $\langle\,\xi s| \xi'\rangle$
(as can be read off from Eqs.(\ref{trans-ampl}) and
(\ref{kg-prop})), we then obtain the expression
\begin{equation}\label{eff-act-exact}
\begin{array}{l}
\Gamma^{(1\!)}\!({\mathcal{B}})\!=\!\frac{i}{2}\!{\displaystyle\int}_{0}^{\infty}\!
\frac{ds}{s}\,e^{-is(m^{2}\!-i\epsilon
)}\int\!d^{4}\xi\,\lim_{\xi'\rightarrow\xi}\left\{\delta(\xi_{2}-\xi_{2}')4\cos({\mathcal{B}}s)
{}^{\displaystyle{\frac{}{}}}\right.\\*[0.3cm]
\hspace*{2cm}\times\left.\frac{1}{4\pi
s}e^{-i\frac{(\xi_{0}-\xi_{0}')^{2}-(\xi_{3}-\xi_{3}')^{2}}{4s}}\sqrt{\frac{1}{2\pi
i\sin(2{\mathcal{B}}s)}}e^{\frac{i}{2}\frac{(\xi_{1}^{2}+\xi_{1}'^{2})\cos(2{\mathcal{B}}s)
-2\xi_{1}\xi_{1}'}{\sin(2{\mathcal{B}}s)}}\right\}\\*[0.3cm]
\hspace*{1cm}=\frac{i}{2\pi}{\displaystyle\int}_{0}^{\infty}\frac{ds}{s^{2}}e^{-is(m^{2}\!-i\epsilon
)}\!\sqrt{\frac{\cot({\mathcal{B}}s)} {4\pi
i}}\!{\displaystyle{\int}}_{-\infty}^{\infty}\!\!d\xi_{1}e^{-i\xi_{1}^{2}\tan({\mathcal{B}}s)}
\!{\displaystyle\int} dtd\xi_{2} d\xi_{3}
\left({\displaystyle\lim_{\xi_{2}'\!\rightarrow\xi_{2}}}\delta(\xi_{2}\!-\!\xi_{2}')\!\right),
\end{array}
\end{equation}
where we have used the result ${ \rm tr}\,
e^{-is{\mathcal{B}}\Sigma_{3}}=4\cos({\mathcal{B}}s)$.

The expression (\ref{eff-act-exact}) contains certain factors for
which we must provide suitable interpretation. Here it is useful
to imagine that our system is defined in a large, but finite,
space-time volume $L^{3}T$. Then, as regards the somewhat
unconventional-looking factor appearing at the end of the
expression in Eq.(\ref{eff-act-exact}), we may give the following
interpretation:
\begin{equation}\label{volume-factor}
\int
dtd\xi_{2}d\xi_{3}\,\lim_{\xi_{2}'\rightarrow\xi_{2}}\delta(\xi_{2}'-\xi_{2})
=\frac{{\mathcal{B}}}{2\pi}TL^{3}.
\end{equation}
This follows once, if one has that
\begin{eqnarray}
&&\int dtd\xi_{3}\sim TL,\label{t&z}\\
&&\lim_{\xi_{2}'\rightarrow\xi_{2}}\delta(\xi_{2}-\xi_{2}')
\sim\frac{\sqrt{{\mathcal{B}}}}{2\pi}L,\label{delta}\\
&&\int d\xi_{2}\sim\sqrt{{\mathcal{B}}}L.\label{obl-vol}
\end{eqnarray}
The relation (\ref{t&z}) is a usual one; we will argue for
Eqs.(\ref{delta}) and (\ref{obl-vol}) below.

First of all, we can infer from Eqs.(\ref{orth-complete}) and
(\ref{landau-kernel}) that
\begin{eqnarray}
\lim_{(\xi_{1}',\xi_{2}')\rightarrow(\xi_{1},\xi_{2})}\delta(\xi_{1}\!-\!\xi_{1}')\delta(\xi_{2}\!-\!\xi_{2}')
\sim\int dxdy\, \frac{{\mathcal{B}}}{4\pi^{2}}\nonumber\\
\sim\left(\frac{\sqrt{{\mathcal{B}}}}{2\pi}L\right)^{2}\hspace{0.15cm}.\label{IR}
\end{eqnarray}
On the other hand, we expect that
${\displaystyle\lim_{\xi_{2}'\rightarrow\xi_{2}}}\delta(\xi_{2}'\!-\!\xi_{2}\!)$
be equal to
${\displaystyle\lim_{\xi_{1}'\rightarrow\xi_{1}}}\delta(\xi_{1}'\!-\!\xi_{1}\!)$;
this must be so since the exchange between $x$ and $y$ gives rise
to the exchange between $\xi_{1}$ and $\xi_{2}$ (while the
variables x and y are to be treated symmetrically). So, based on
Eq.(\ref{IR}), we have Eq.(\ref{delta}). To have
Eq.(\ref{obl-vol}) justified, note that
\begin{equation}\label{delta-vol}
\lim_{\eta_{2}'\rightarrow\eta_{2}}\delta(\eta_{2}\!-\!\eta_{2}')
\sim\frac{1}{2\pi}\int d\xi_{2}\,,
\end{equation}
which is the usual relation when two mutually conjugate variables
are involved. Then, following the same step that we have used to
derive Eq.(\ref{delta}) above, it is also possible to show that
\begin{equation}\label{obl-conj}
\lim_{\eta_{2}'\rightarrow\eta_{2}}\delta(\eta_{2}-\eta_{2}')
\sim\frac{\sqrt{{\mathcal{B}}}}{2\pi}L\,.
\end{equation}
[Here, instead of Eq.(\ref{landau-kernel}), one could utilize the
expression $\langle\vec{r}|\vec{\eta} \rangle =
\frac{\sqrt{{\mathcal{B}}}}{2\pi}\,
e^{i\sqrt{{\mathcal{B}}}(x\eta_{1}+y\eta_{2})
+i\frac{{\mathcal{B}}}{2}xy+i\eta_{1}\eta_{2}}$\ ]. Based on
Eqs.(\ref{delta-vol}) and (\ref{obl-conj}), we have
Eq.(\ref{obl-vol}).

With Eq(\ref{volume-factor}) used in the expression
(\ref{eff-act-exact}), we may now write
$\Gamma^{(\!1\!)}\!({\mathcal{B}})={\mathcal{L}}^{(\! 1 \!)}
\!({\mathcal{B}}) L^{3} T$, ${\mathcal{L}}^{(\!
1\!)}\!({\mathcal{B}})$ giving the one-loop contribution to the
effective Lagrangian density. The $\xi_{1}$-integration in
Eq.(\ref{eff-act-exact}) is readily performed, but the remaining
proper-time integration is plagued with divergence at the $s=0$
end. The amplitude needs to be renormalized. Explicitly, from
Eq.(\ref{eff-act-exact}) (after the $\xi_{1}$-integration), we
have
\begin{equation}\label{eff-lag}
\begin{array}{l}
{\mathcal{L}}^{(\! 1\!)}\!({\mathcal{B}}) = \frac{{\mathcal{B}}}{
8 \pi^{2}} {\displaystyle\int_{0}^{\infty}} \frac{ds}{s^{2}}\,
e^{-im^{2}s}\cot({\mathcal{B}}s)\\*[0.3cm] \hspace*{1.25cm} =
({\mathcal{B}}\!-\!independent\ const.) -\frac{1}{24\pi^{2}}
{\displaystyle\int_{0}^{\infty}} \frac{ds}{s}\,
e^{-im^{2}s}{\mathcal{B}}^{2}\\*[0.3cm] \hspace*{1.7cm} +\left\{
\frac{1}{8\pi^{2}} {\displaystyle\int_{0}^{\infty}}
\frac{ds}{s^{3}} \, e^{-im^{2}s} \left[
{\mathcal{B}}s\cot({\mathcal{B}}s)-1
+\frac{1}{3}({\mathcal{B}}s)^{2} \right]\right\}\ .
\end{array}
\end{equation}
The divergent contribution is contained in the first two terms of
the second form in Eq.(\ref{eff-lag}); the expression appearing
inside the curly bracket is finite. The
${\mathcal{B}}$-independent constant in Eq.(\ref{eff-lag}) is
insignificant and can be dropped, while the second term, which is
proportional to ${\mathcal{B}}^{2}$ and has a logarithmically
divergent coefficient, gets cancelled as one introduces the
renormalization counterterm. In fact the above expression for
$\mathcal{L}^{(\!1\!)}\!({\mathcal{B}})$, which we found with the
help of the oblique phase-space formulation, is in complete
agreement with the result given in Refs.[11-13]. But, according to
our discussion, it is evident that two oblique phase-space
coordinates $\xi_{1}$ and $\xi_{2}$ assume very different role
with respect to ultraviolet renormalization (and with respect to
the contribution to the space-time volume factor also). This is in
contrast with the roles assumed by coordinates x and y in the
usual position-space formulation.
\\

\section{INTERACTION}

%\addtocounter{chapter}{1}\setcounter{equation}{0}

In the previous sections, we have seen that the quadratic
Lagrangian of the charged matter field with a uniform background
magnetic field simplifies greatly in oblique variables. The matter
field propagator can be obtained easily and used to calculate the
effective action for the background magnetic charge. Basically,
the `free' Lagrangian describes the kinematics of charged
particles in each Landau level.

There can be interaction between charged particles in different
Landau levels. The interaction Lagranian of the matter field,
including the external potential depending on space, was studied
in oblique variables in Sec.~II. It shows how the interaction term
can be expressed as the interaction between particles belonging to
different Landau levels.  When the dynamical electromagnetic field
is coupled the matter field, one has to be careful as photons does
not see the magnetic field.  While we may want to express the
matter field as a function of the oblique varibles, we want to
keep the gauge field as a function of ordinary space time
coordinates or their conjugate momentum variables. In this
section, let us focus on the interaction term between the
electromagnetic field and the Dirac field for the simplicity.

The Maxwell action for the dynamics electromagntic field is
\begin{equation}
S_{M}=-\frac{1}{4}\int d^{4}x F_{\mu\nu}(x)F^{\mu\nu}(x).
\end{equation}
The gauge interaction of the Dirac field to the gauge field is
\begin{equation}\label{inter}
S_{1}=\int d^{4} x \bar{\Psi}(x)\gamma_{\mu}\Psi(x)A_{\mu}(x) .
\end{equation}
We may introduce the gauge fixing condition, like the Feynmann
gauge-fixing term $-\frac{1}{2}(\partial\cdot A(x))^{2}$, when we
are working on the perturbative expansion. We change the arguments
$x$, $y$ for the Dirac field into the oblique phase space
variables $\xi_{1}$ and $\xi_{2}$ as before:
\begin{equation}
\Psi(x)=\int d^{2}\xi
\langle\vec{x}|\vec{\xi}\rangle\Phi(\vec{\xi},z,t).
\end{equation}
But for the Maxwell field, it is the momentum
$\vec{q}=(q_{1},q_{2})$ which naturally diagonalizes the quadratic
action, so we choose the momenta to describe the Maxwell field,
\begin{equation}
A_{\mu}(x)=\int d^{2}\vec{q}
e^{i\vec{q}\cdot\vec{x}}a_{\mu}(\vec{q},z,t)
\end{equation}
where there are 4 modes $a_{\mu}(\vec{q},z,t)$ for a momentum
$\vec{q}$, and
$a_{\mu}(\vec{q},z,t)=a^{\dag}_{\mu}(-\vec{q},z,t)$. In terms of
these fields, the interaction (\ref{inter}) is rewritten as
\begin{equation}
\int dt\ dz\ \int d^{2}q\ d^{2}\xi d^{2}\xi^{\prime}
C(\vec{\xi},\vec{\xi}';\vec{q})\bar{\Phi}(\xi,z,t)\gamma^{\mu}
\Phi(\vec{\xi}',z,t)a_{\mu}(\vec{q},z,t) ,
\end{equation}
where $C(\vec{\xi},\vec{\xi}';\vec{q})=\int d^{2}x
\langle\vec{\xi}|\vec{x}\rangle e^{i\vec{q}\cdot\vec{x}}
\langle\vec{x}|\vec{\xi'}\rangle$. It may be more illuminating to
describe the Dirac field in terms of the Landau level indices
instead of $\xi$ as in the previous sections.  Then the above
gauge interaction becomes
\begin{equation}\label{landau-int}
\begin{array}{c}{\displaystyle{
\int\,dtdz\,\int d^{2}q\ d\xi_{2} d\xi^{\prime}_{2}
{\mathcal{C}}_{mn}(\xi_{2},\xi_{2}';\vec{q})\bar{\Phi}_{m}(\xi_{2},z,t)\left
\{ \left[\cos\frac{\theta_{m} -\theta_{n}}{2}-\sin\frac{\theta_{m}
-\theta_{n}}{2} \gamma^{2}\right]\right. }}
\\{\displaystyle{
\left. \times\gamma^{\mu}
a_{\mu}(\vec{q},z,t)+2\sin\frac{\theta_{n}}{2}\left [
\cos\frac{\theta_{m}}{2}
-\sin\frac{\theta_{m}}{2}\gamma^{2}\right]
a_{2}(\vec{q},z,t)\right \}\Phi_{n}(\xi'_{2},z,t),}}
\end{array}
\end{equation}
where $\theta_{n}$, $\Phi_{n}(\xi_{2},z,t)$ are defined by
Eqs.(\ref{dirac-obl-landau-transf}),(\ref{landau-eff-transf}) and
(\ref{eff-transf-matrix}), and the kernel ${\mathcal{C}}_{mn}$ is
defined as follows(the form of
${\mathcal{W}}^{(n)}(\vec{x};\xi_{2})$ is given as
Eq.(\ref{pos-landau-kernel})):
\begin{equation}
\int d^{2}x {\mathcal{W}}^{(m)}(\vec{x};\xi_{2})
e^{i\vec{q}\cdot\vec{x}}{\mathcal{W}}^{(m)}(\vec{x};\xi_{2}').
\end{equation}
This kernel gives the strength with which the transition between
two Landau levels occur in the presence of dynamic Maxwell field.
It is  explicitly calculated to be ($n>m$ assumed):
\begin{equation}\label{MFkernel}
\begin{array}{c}{\displaystyle{
{\mathcal{C}}_{mn}=i^{n-m}\delta[\xi_{2}-\xi_{2}'-\frac{q_{1}}{\sqrt{{\mathcal{B}}}}]
e^{i\frac{q_{1}q_{2}}{{\mathcal{B}}}
+i\frac{q_{2}(\xi_{2}+\xi_{2}')}{2{\mathcal{B}}}
-\frac{q_{2}^{2}}{4{\mathcal{B}}}}
(\frac{q_{1}+iq_{2}}{\sqrt{2{\mathcal{B}}}})^{n-m} }}\\
{\displaystyle{ \times\sqrt{\frac{n!}{m!(n-m)!}}
M(-m,n-m+1,(\frac{q_{1}+iq_{2}}{\sqrt{2{\mathcal{B}}}})
(\frac{iq_{2}}{\sqrt{2{\mathcal{B}}}})), }}
\end{array}
\end{equation}
where $M(-m,n-m+1,A)$ in the expression is just a polynomial of
$A$, but takes the form of confluent hypergeometric function (in
this case the Kummer function):
\begin{equation}
M(a,b,A)=\sum_{l=0}^{\infty}\frac{(a)_{l}}{(b)_{l}l!}A^{l}.
\end{equation}

The kernel (\ref{MFkernel}) and the interaction (\ref{landau-int})
show some remarkable behaviors. First of all, the delta function
factor in the kernel gives a relation which somewhat resembles the
momentum conservation in x direction. Recall that the definition
of $\xi_{2}$ is
$\frac{1}{\sqrt{{\mathcal{B}}}}(p_{x}+\frac{1}{2}{\mathcal{B}}y)$,
so in the weak field limit the delta function factor precisely
gives the momentum conservation.

The interaction represented in oblique phase-space description may
seem somewhat complicated, but it is because the quadratic part of
the theory only respects rotation and boost symmetry in z
direction. The boost symmetry along the $z$ direction still exists
in the expressions (\ref{landau-int}) and (\ref{MFkernel}). The
rotation symmetry is not manifest in the formula, but it comes
from our choice of $\xi_{2}$ to label the degeneracy of Landau
levels. If we are only interested in the interaction including the
lowest Landau level, the interaction looks simpler. The term with
$n=0$ looks like $a_{\mu}\bar{\Phi}_{m}[\cos\frac{\theta_{m}}{2}
-\sin\frac{\theta_{m}}{2}\gamma^{2}]\gamma^{\mu}\Phi_{0}$.
Interaction within the lowest Landau levels, which is the most
interesting part in some problems, simply takes the form of
$a_{\mu}\bar{\Phi}_{0}\gamma^{\mu}\Phi_{0}$.

Furthermore, if the external magnetic field ${\mathcal{B}}$ is
large, the kernel (\ref{MFkernel}) shows that the transition
between two different Landau levels are supressed by the factor
$(\frac{q_{1}+iq_{2}}{\sqrt{2{\mathcal{B}}}})^{\Delta}$ where
$\Delta$ is the difference of two Landau level indices. For
example, if we integrate out the first Landau level effect, the
effective interaction of the lowest Landau level and Maxwell
fields will contain a term of order
$(\frac{q_{1}^{2}+q_{2}^{2}}{2{\mathcal{B}}})$.

\section{CONCLUSION}

In studying nonrelativistic or relativistic field theory systems
in a background magnetic field, we have exhibited the advantage of
using quantum field operators defined in the oblique phase space.
By the very coordinate choice the mode rearrangements related to
the appearance of Landau levels are incorporated in a natural
manner.  In addition, we have expressed the interaction Lagrangian
in oblique variables so that Landau level indices appear. Also the
phenomenon of dimensional reduction in a strong background
magnetic field becomes evident in our oblique phase-space
description.

Thus the investigation of the interacting field theory models in
our formalism should give us valuable insights as regards to many
interesting physical characteristics exhibited by the physical
systems in a strong magnetic field. Our formalism could be a
starting point for the standard field theoretic development,
including the discussions on higher-order loop effects. Although
much work has been done already in this direction[3-6,9,10], it is
hoped that the oblique phase-space field description may serve a
useful purpose in uncovering yet unknown aspects. We intend to
report on such consideration in our future work.

\centerline{\bf ACKNOWLEDGMENTS}\vspace{0.5cm}

This work was supported in part by BK21 project of the Ministry of
Education, Korea (S.K. and C.L.), Korea Research Foundation Grant
2001-015-DP0085 (C.L.), and KOSEF 1998 Interdisciplinary Research
Grant 98-07-02-07-01-5 (K.L.).

\newpage

\centerline{\bf APPENDIX A}\vspace{0.5cm}
\setcounter{section}{1}\setcounter{equation}{0}
\renewcommand{\theequation}{\Alph{section}.\arabic{equation}}

Here we will derive the explicit form of the kernel
$\langle\vec{r}|\vec{\xi}\rangle$, satisfying the conditions
(\ref{orth-complete}), (\ref{xi-repre}) and (\ref{eta-repre}). The
matrices $C$ and $D$ will be assumed to be nonsingular. Then,
$D^{-1}C$ is a symmetric matrix (thanks to the condition
$CD^{T}=DC^{T}$ in Eq.(\ref{simplectic})) and we may well cast
Eq.(\ref{xi-repre}) in the form
\begin{equation}\label{xi-eq}
\frac{\partial}{\partial
x^{i}}\langle\vec{r}|\vec{\xi}\rangle=i\left[\sqrt{2}(D^{-1})_{ij}\xi^{j}
-(D^{\!-1}\!C)_{ij}x^{j}\right]\langle\vec{r}|\vec{\xi}\rangle.
\end{equation}
This has the general solution
\begin{equation}\label{xi-eq-sol}
\langle\vec{r}|\vec{\xi}\rangle=H(\vec{\xi})e^{i[\sqrt{2}x^{i}(D^{\!-1})_{ij}\xi^{j}
-\frac{1}{2}x^{i}(D^{\!-1}\!C)_{ij}x^{j}]},
\end{equation}
$H(\vec{\xi})$ being any function of $\vec{\xi}$ at this stage. If
the normalization condition (\ref{orth-complete}) is imposed with
the form (\ref{xi-eq-sol}), $H(\vec{\xi})$ is determined up to an
arbitrary phase, that is,
\begin{equation}\label{normalise}
H(\vec{\xi})=\left|(2\pi)^{d}\det(\frac{D}{\sqrt{2}})\right|^{-\frac{1}{2}}e^{i\gamma(\vec{\xi})},
\end{equation}
and so
\begin{equation}\label{up-to-phase-sol}
\langle\vec{r}|\vec{\xi}\rangle= \frac{1}{(2\pi)^{d/2}\sqrt{|
\det(\frac{D}{\sqrt{2}})|}}
e^{i\gamma(\vec{\xi})}e^{i[\sqrt{2}x^{i}(D^{\!-1})_{ij}\xi^{j}
-\frac{1}{2}x^{i}(D^{\!-1}\!C)_{ij}x^{j}]}.
\end{equation}

The condition (\ref{eta-repre}) may be used to fix the phase
$\gamma(\vec{\xi})$ in Eq.(\ref{up-to-phase-sol}). Here, for
$\xi^{i}$ given as in Eq.(\ref{g-ob-pos}), we note that the
conjugate momentum variable $\eta^{i}$ (see Eq.(\ref{g-ob-mom}))
should have the general form
\begin{equation}\label{general-eta}
\eta^{i}=\frac{1}{\sqrt{2}}\left[-(D^{T})^{-1}_{ij}x^{j}+(C^{T})^{-1}_{ij}p^{j}\right]
+(\Delta)_{ij}\xi^{j},
\end{equation}
where $\Delta$ is an arbitrary symmetric matrix. [According to
this, the matrices $E$ and $F$, obeying the restrictions in
Eq.(\ref{simplectic}), can be expressed as $E=-(D^{T})^{-1}+\Delta
C$, $F=(C^{T})^{-1}+\Delta D$]. Using the form
(\ref{general-eta}), the condition (\ref{eta-repre}) now reads
\begin{equation}\label{eta-eq}
i\frac{\partial}{\partial\xi^{i}}\langle\vec{r}|\vec{\xi}\rangle
=\frac{1}{\sqrt{2}}\left[-(D^{T})^{-1}_{ij}x^{j}-i(C^{T})^{-1}_{ij}\frac{\partial}{\partial
x^{j}}\right]\langle\vec{r}|\vec{\xi}\rangle
+(\Delta)_{ij}\xi^{j}\langle\vec{r}|\vec{\xi}\rangle,
\end{equation}
and this in turn implies the following equation for
$\langle\vec{r}|\vec{\xi}\rangle$ (given by
Eq.(\ref{up-to-phase-sol})):
\begin{equation}\label{eta-eq-sol}
\frac{\partial\gamma(\vec{\xi})}{\partial\xi^{i}}\langle\vec{r}|\vec{\xi}\rangle
=-\left[(CD^{T})^{-1}_{ij}\xi^{j}+(\Delta)_{ij}\xi^{j}\right]
\langle\vec{r}|\vec{\xi}\rangle\ .
\end{equation}
To obtain Eq.(\ref{eta-eq-sol}), we have made use of
Eq.(\ref{xi-repre}) and also the relation $CD^{T}=DC^{T}$.
Integrating Eq.(\ref{eta-eq-sol}), we find
\begin{equation}\label{phase}
\gamma(\vec{\xi})=-\frac{1}{2}\xi^{i}\left((CD^{T})^{-1}+\Delta\right)_{ij}\xi^{j}+const.
\end{equation}
Without loss of generality the arbitrary constant in
Eq.(\ref{phase}) may be set to zero. Then we end up with the
following expression for $\langle\vec{r}|\vec{\xi}\rangle$:
\begin{equation}\label{kernel-sol}
\langle\vec{r}|\vec{\xi}\rangle= \frac{1}{(2\pi)^{d/2}\sqrt{|
\det(\frac{D}{\sqrt{2}})|}}
e^{-\frac{i}{2}[x^{i}(D^{\!-1}\!C)_{ij}x^{j}+\xi^{i}((CD^{T})^{\!-1}\!+\Delta)_{ij}\xi^{j}]}
e^{i\sqrt{2}x^{i}(D^{\!-1}\!)_{ij}\xi^{j}}.
\end{equation}
With the choice $\Delta=0$ made, this reduces to the expression in
Eq.(\ref{kernel}).
\\

\centerline{\bf APPENDIX B}\vspace{0.5cm}
\setcounter{section}{2}\setcounter{equation}{0}

For the given background magnetic field $\vec{B}=B_{0}\hat{z}$,
another popular choice for the vector potential is the expression
in the Landau gauge, $\vec{A}(\vec{r})=(0,B_{0}x,0)$. Then we
have, instead of the quadratic Lagrangian in
Eq.(\ref{sch-pos-lag}), the expression (with ${\mathcal{B}}\equiv
-qB_{0}>0$)
\begin{equation}\label{coul-gauge-lag}
\int
d^{3}\vec{r}\Psi^{\dag}(\vec{r},t)\left[i\frac{\partial}{\partial
t}-\frac{1}{2m}(-i\partial_x)^{2}\!-\!\frac{1}{2m}(-i\partial_{y}+{\mathcal{B}}x)^{2}
\!-\!\frac{1}{2m}(-i\partial_{z})^{2}\right]\Psi(\vec{r},t).
\end{equation}
In this case, the following set of variables may be considered
instead of $(x,y,p_{x},p_{y})$:
\begin{equation}\label{coul-obl-transf}
\begin{array}{ll}
{\displaystyle\xi_{1}=\frac{1}{\sqrt{{\mathcal{B}}}}(p_{y}+{\mathcal{B}}x),}&
{\displaystyle\xi_{2}=\frac{1}{\sqrt{{\mathcal{B}}}}(p_{x}+{\mathcal{B}}y),}\\*[0.4cm]
{\displaystyle\eta_{1}=\frac{1}{\sqrt{{\mathcal{B}}}}p_{x},}&
{\displaystyle\eta_{2}=\frac{1}{\sqrt{{\mathcal{B}}}}p_{y}.}
\end{array}
\end{equation}
With this choice and the corresponding field
$\Phi(\xi_{1},\xi_{2},\xi_{3}\!\equiv\!z,t)$, we can transform the
form (\ref{coul-gauge-lag}) again to the expression given in
Eq.(\ref{sch-obl-lag}). So, only if we have the explicit field
transformation, i.e., the appropriate expression for the kernel
$\langle\vec{r}|\vec{\xi}\rangle$ in Eq.(\ref{g-fourier}), our
symmetric-gauge-based discussion should carry over to the Landau
gauge case.

We may use the result of Appendix A to find the Landau-gauge
kernel $\langle\vec{r}|\vec{\xi}\rangle$ explicitly. In this case,
the symmetric matrix $\Delta$ in Eq.(\ref{general-eta}) is chosen
as
\begin{equation}\label{coul-delta}
\Delta=(CD^{T})^{-1},
\end{equation}
so that
$(\Delta)_{ij}\xi^{j}=\frac{1}{\sqrt{2}}\left[(D^{T})^{-1}_{ij}x^{j}+(C^{T})^{-1}_{ij}p^{j}\right]$,
and hence the conjugate momentum variables may read
\begin{equation}\label{general-coul-eta}
\eta^{i}=\frac{1}{\sqrt{2}}2(C^{T})^{\!-\!1}_{ij}p^{j}.
\end{equation}
Comparing the relationships in Eq.(\ref{coul-obl-transf}) with
those given by Eqs.(\ref{g-ob-mom}) and (\ref{general-coul-eta}),
the matrices C, D and $\Delta$ are found immediately:
\begin{equation}
C=\left(\begin{array}{cc}\sqrt{2{\mathcal{B}}}&0\\
0&\sqrt{2{\mathcal{B}}}\end{array}\right)\hspace{0.1cm},\hspace{0.15cm}
D=\left(\!\begin{array}{cc}\!\sqrt{\frac{2}{{\mathcal{B}}}}\!&0\\
0&\!\sqrt{\frac{2}{{\mathcal{B}}}}\!\end{array}\!\right)\hspace{0.1cm},\hspace{0.15cm}
\Delta=\frac{1}{2}\left(\begin{array}{cc}1&0\\0&1\end{array}\right)\hspace{0.1cm}.
\end{equation}
Hence, from Eq.(\ref{kernel-sol}), the desired kernel is found to
be
\begin{equation}
\begin{array}{l}
\langle\vec{r}|\vec{\xi}\rangle=\frac{\sqrt{{\mathcal{B}}}}{2\pi}
e^{-i({\mathcal{B}}xy+\xi_{1}\xi_{2})}e^{i{\mathcal{B}}(x\xi_{2}+y\xi_{1})}
\delta(\xi_{3}-z)\\*[0.3cm]
\hspace*{0.8cm}=\frac{\sqrt{{\mathcal{B}}}}{2\pi}e^{-i(\mathcal{\sqrt{B}}x-\xi_{1})(\mathcal{\sqrt{B}}y-\xi_{2})}
\delta(\xi_{3}-z).
\end{array}
\end{equation}
We remark that analogous discussions can be given for relativistic
field systems also.
\\


\newpage

\begin{thebibliography}{99}
\bibitem{1}L. D. Landau, Z. Phys. {\bf 64}, 629 (1930); L.D.Landau and
E.M.Lifshitz, \textit{Quantum Mechanics (Nonrelativistic Theory)}
(Pergamon Press,1959), p.456.

\bibitem{2}See, for instance, R.
Prange and S. Girvin (editors), \textit{The Quantum Hall Effect},
2nd ed. (Springer, Berlin, 1990); M. Stone, \textit{Quantum Hall
Effect} (World Scientific, Singapore, 1992).

\bibitem{3}V. P.
Gusynin, V. A. Miransky, and I. A. Shovkovy, Phys. Rev. Lett. {\bf
73}, 3499 (1994); Phys. Rev. {\bf D52}, 4718 (1995); Phys. Lett.
{\bf B349}, 477 (1995); Phys. Rev. {\bf D52}, 4747 (1995); Nucl.
Phys. {\bf B462}, 249 (1996); Phys. Rev. Lett. {\bf 83}, 1291
(1999).
\bibitem{4}C. N. Leung, Y. J. Ng and A. W. Ackley, Phys. Rev.
{D54}, 4181 (1996); D. -S. Lee, C. N. Leung and Y. J. Ng, ibid.
{\bf 55}, 6504 (1997).
\bibitem{5}D. K. Hong, Y. Kim and S. -J. Sin, Phys.
Rev. {\bf D54}, 7879 (1996); D. K. Hong, ibid. {\bf 57}, 3759
(1998).
\bibitem{6}G. W. Semenoff, I. A. Shovkovy and L. C. R. Wijewardhana, Phys. Rev.
{\bf D60}, 105024 (1999).
\bibitem{7}S. M. Girvin and T. Jach, Phys. Rev. {\bf B29}, 5617 (1984).
\bibitem{8}G. Dunne, R. Jackiw and C. Trugenberger, Phys. Rev. {\bf D41},
661 (1990); G. Dunne and R. Jackiw, Nucl. Phys. Proc. Suppl. {\bf
33C}, 114 (1993).
\bibitem{9}S. Iso, D. Karabali and B. Sakita, Nucl. Phys.
{\bf B388}, 700 (1992); Phys. Lett. {B 296}, 143 (1992).
\bibitem{10}B.
Sakita, Phys. Lett. {\bf B315}, 124 (1993); R. Ray and B. Sakita,
Ann. Phys. {\bf 230}, 131 (1994); K. Ishikawa, N. Maeda, T. Ochiai
and H. Susuki, Phys. Rev. {\bf B58}, 1008 (1998).
\bibitem{11}W. Heisenberg and H. Euler,
Z. Phys. {\bf 98}, 714 (1936); V. Weisskopf, K. Dan. Vidensk.
Selsk. Mat. Fys. Medd. {\bf 14}, 6 (1936).
\bibitem{12}J. Schwinger, Phys. Rev. {\bf 82}, 664 (1951).
\bibitem{13}C. Itzykson and J. -B. Zuber, \textit{Quantum Field Theory}
(McGraw-Hill, 1980), p.100 and p.195; W. Dittrich and H. Gies,
``Probing the Quantum Vacuum: Perturbative Effective Action
Approach in Quantum Electrodynamics and its Application''
(Springer-Verlag, 2000).
\bibitem{14}B. Ovrut and H.
Schnitzer, Phys. Rev. {\bf D21}, 3369 (1980); S. Weinberg, Phys.
Lett. {\bf 91B}, 51 (1980).
\bibitem{15}H. Georgi, \textit{Weak
Interactions and Modern Paritcle Theory} (Benjamin/Cummings, Menlo
Park, California, 1984), p.124.
\end{thebibliography}
\end{document}